\documentclass[12pt]{article}
\usepackage{graphicx}

\newcommand{\be}{\begin{equation}}
\newcommand{\ee}{\end{equation}}
\newcommand{\bea}{\begin{eqnarray}}
\newcommand{\eea}{\end{eqnarray}}
\newcommand{\al}{\\[0.4cm]}
\title{\Large \bf STUDY OF THE O(N) LINEAR $\sigma$ MODEL AT
FINITE TEMPERATURE USING THE 2PPI EXPANSION}
\author{\normalsize H. VERSCHELDE and J. DE PESSEMIER\thanks
{Research Assistant of the Fund for 
Scientific Research - Flanders (Belgium) (F.W.O.)}\\ 
\it University of Gent\\
\it Department of Mathematical Physics and Astronomy \\
\it Krijgslaan 281-S9 \\
\it B-9000 GENT, BELGIUM} 
\date{}
\begin{document}
\maketitle
\vskip 36pt

\begin{abstract}

We show that a new expansion which sums seagull and bubble
graphs to all orders, can be applied to the O(N)-linear
$\sigma$-model at finite temperature. We prove that this
expansion can be renormalised with the usual counterterms in a
mass independent scheme and that Goldstone's theorem is
satisfied at each order. At the one loop order of this
expansion, the Hartree result for the effective potential (daisy
and superdaisy graphs) is recovered. We show that at one loop
2PPI order, the self energy of the $\sigma$-meson can be
calculated exactly and that diagrams are summed beyond the
Hartree approximation.
\end{abstract}

\newpage

\section{Introduction}

In this paper, we will study the O(N) lineair $\sigma$-model at finite temperature, 
using the 2PPI expansion [1,2]. This model has always been a fertile ground to test
ideas and check approximations in finite temperature quantum field theory [3]
and has recently attracted renewed interest [4-11] because of its relevance to
the thermodynamics of chiral symmetry in Q.C.D. Many treatments of finite T O(N) linear 
$\sigma$-model use the Hartree approximation which sums bubble graphs (daisy and
superdaisy graphs). The standard way of summing these graphs is to use the 2PI
expansion or CJT-method [12]. In this approach, only the 2PI diagrams are retained, which
do not seperate in two pieces when two internal lines are cut. This sums self energy
insertions but comes at the cost of introducing a self consistency condition which in
general entails intractable non-local integral equations. When restricted to order $\lambda$, 
the
2PI expansion sums daisy and superdaisy graphs which alleviates some of the problems
at finite T [4]. However, it is not feasible to extend this approach to higher order
in $\lambda$ and therefore some of the basic problems of the O(N) linear $\sigma$-model
at finite T are still unsolved. Another problem encountered is renormalizability. In
[4], the daisy and superdaisy graphs are summed with the 2PI expansion at O$(\lambda)$, using
bare perturbation theory. It is found that the effective bubble mass is finite when
the coupling constant runs according to a "non-perturbative" $\beta$ function which does
not agree with the perturbative one. For N = $\infty$, these two $\beta$ functions 
coincide and for this reason, many treatments of finite T O(N) linear $\sigma$-model
use the N $\rightarrow \infty$ limit. One can ask oneself if these are genuine
renormalisation problems or just problems due to inconsistent renormalisation.
Finally there is the Goldstone theorem at finite T. Although originally there were some
papers claiming that Goldstone's theorem was violated at finite T, there is now
ample evidence [6,10,13] that it is valid at all temperatures. It would however be
preferable to have a simple all orders proof of this important fact. In this paper, 
we will address these problems using the O(N) linear $\sigma$-model as a simple
model of spontaneous symmetry breaking.

\section{The 2PPI expansion}

The 2PPI expansion is an approximation scheme for calculating the effective
action for local composite operators. It was introduced in [1] for $\lambda \phi^4$
theory with composite operator $\phi^2$. The corresponding effective potential
can be viewed as the minimum of the energy density within the class of wavefunctionals
with fixed expectation values for the elementary fields and one or more local composite
operators. Minimization with respect to the values of the composite operators yields
gapequations which sum infinite series of Feynman diagrams. In the case of the 2PPI
expansion with local composite operators quadratic in the fields, the gap equations
sum bubble graphs, sometimes also called tadpole graphs. The 2PPI expansion is to the
effective action for a local composite operator what the 1PI expansion is to the
ordinary effective action for elementary fields or what the 2PI expansion (or CJT
formalism [12]) is to the effective action for a bilocal composite operator. In this
section, we will derive the 2PPI expansion for the $O(N)$ linear $\sigma$-model with
composite operators $\phi_i(x)\phi_j(x)$. Our derivation will not use the formalism
of effective actions and Legendre transforms [14], but will be more directly based on
Feynman diagram analysis. This will enable a transparent proof of renormalizability
as one of us has shown for $\lambda \phi^4$ [2].
\al 
The Lagrangian for the $O(N)$ linear $\sigma$-model reads
\bea
\cal{L} & = & \frac{1}{2} \partial_{\mu} \phi_i \partial_{\mu}\phi_j 
              + \frac{m^2_{ij}}{2} \phi_i \phi_j + \frac{\lambda}{8} (\phi_{ii})^2 
              + \delta \cal{L}  \\ \nonumber
        & = & \frac{1}{2} \partial_{\mu} \phi_i \partial_{\mu} \phi_j 
              + \frac{m^2_{ij}}{2} \phi_i \phi_j 
              + \frac{\lambda_{ijk\ell}}{4!} \phi_i \phi_j \phi_k \phi_{\ell}
              + \delta \cal{L}
\eea
with
\be
\lambda_{ijk\ell} = \lambda (\delta_{ij} \delta_{k\ell} + \delta_{ik} \delta_{j\ell}
+ \delta_{i\ell} \delta_{jk})
\ee
In this section, we will treat the unrenormalised 2PPI expansion and hence neglect
all contributions from the counterterm Lagrangian. The way to get to the 2PPI
expansion is to start from the 1PI expansion and sum all the seagull and bubble
graphs. These insertions arise in 2PPR or two particle point reducible graphs because
they disconnect from the rest of the diagram where two lines meeting at the same
point (the 2PPR-point) are cut (fig.~1). We notice that seagull and bubble graphs
contribute to the self energy as effective mass terms proportional to $\varphi_i \varphi_j$
and $\Delta_{ij} = \langle \phi_i \phi_j \rangle_c$ respectively. A short diagrammatical
analysis suggests that all 2PPR insertions can be summed by simply deleting the 2PPR
graphs from the 1PI expansion and introducing the effective mass :
\be
\overline{m}^2_{ij} = m^2_{ij} + \lambda [\varphi_i \varphi_j + \Delta_{ij}]
+ \frac{\lambda}{2} [\varphi^2 + \Delta_{kk}] \delta_{ij}
\ee
in the remaining 2PPI graphs. This is too naive though since there is a double
counting problem which can be easily understood in the simple case of the 2 loop
vacuum diagram (daisy graph with two petals) of fig. 2.a. Each petal can be seen
as a selfenergy insertion in the other, so there is no way of distinguishing one
or the other as the remaining 2PPI part. The trick which solves this combinatorial 
problem is to earmark one of the petals by applying a derivative with respect to
$\varphi_k$ (fig. 2.b). This fixes the 2PPI remainder (which contains the earmark) in
a unique way. Now, there are two ways in which the derivative can hit a $\varphi$ field.
It can hit an explicit $\varphi$ field which is not a wing of a seagull or it can 
hit a wing of a seagull or implicit $\varphi$ field hidden in the effective mass.
We therefore have
\be
\frac{\delta}{\delta \varphi_k} \Gamma^{1PI}_q (m^2,\varphi) = \frac{\partial}
{\partial \varphi_k} \Gamma^{2PPI}_q (\overline{m}^2,\varphi) + \left[
\lambda \varphi_k \delta_{ij} + \lambda (\delta_{ik} \varphi_j + \delta_{jk} \varphi_i)
\right] \frac{\partial \Gamma^{2PPI}_q}{\partial \overline{m}^2_{ij}} (\overline{m}^2,
\varphi)
\ee
where $\Gamma^{1PI} = S(\varphi) + \Gamma^{1PI}_q$ or using the equation for the
effective mass :
\be
\frac{\delta}{\delta \varphi_k} \Gamma^{1PI}_q (m^2,\varphi) = \frac{\delta}{\delta
\varphi_k} \Gamma^{2PPI}_q (\overline{m}^2,\varphi) - \left[ \lambda \frac
{\delta \Delta_{ij}}{\delta \varphi_k} + \frac{\lambda}{2} \delta_{ij} \frac
{\delta \Delta_{\ell\ell}}{\delta \varphi_k}\right] \frac{\partial \Gamma^{2PPI}_q}
{\partial \overline{m}^2_{ij}} (\overline{m}^2,\varphi)
\ee
Using the same type of combinatorial argument, we have :
\be
\frac{\partial \Gamma^{1PI}_q}{\partial m^2_{ij}} (m^2,\varphi) = \frac
{\partial \Gamma^{2PPI}_q}{\partial \overline{m}^2_{ij}} (\overline{m}^2,\varphi)
\ee
and since
\be
\frac{\Delta_{ij}}{2} = \frac{\partial \Gamma^{1PI}_q}{\partial m^2_{ij}} (m^2,\varphi)
\ee
we find the following gap equation for $\Delta_{ij}$ :
\be
\frac{\Delta_{ij}}{2} = \frac{\partial \Gamma^{2PPI}_q}{\partial \overline{m}^2_{ij}} 
(\overline{m}^2,\varphi)
\ee
The gap equation (8) can be used to integrate (5) and we finally obtain
\be
\Gamma^{1PI}(\overline{m}^2,\varphi) = S(\varphi) + \Gamma^{2PPI}_q (\overline{m}^2,
\varphi) - \frac{\lambda}{8} \int d^D x \left( (\Delta_{ii})^2 + 2(\Delta_{ij})^2\right)
\ee
This equation gives the 1PI effective action in terms of the 2PPI effective action and a
term which corrects for double counting. The 2PPI effective action is just the 1PI
effective action without 2PPR graphs and with the effective mass given by equation (3)
running in the internal lines.
For $m^2_{ij} = \delta_{ij} m^2$ we can make use of $O(N)$ symmetry to define :
\be
\overline{m}^2_{ij} = \frac{\varphi_i \varphi_j}{\varphi^2} \overline{m}^2_{\sigma} + 
\left(\delta_{ij} - \frac{\varphi_i \varphi_j}{\varphi^2}\right)
\overline{m}^2_{\pi}
\ee
and
\be
\Delta_{ij} = \frac{\varphi_i \varphi_j}{\varphi^2} \Delta_{\sigma} + \left(
\delta_{ij} - \frac{\varphi_i \varphi_j}{\varphi^2}\right) \Delta_{\pi}
\ee
so that the equation for the effective masses can be written as :
\bea
\overline{m}^2_{\sigma} & = & m^2 + \frac{3\lambda}{2} \left[ \varphi^2 + \Delta_{\sigma}
+ \frac{N-1}{3} \Delta_{\pi}\right] \nonumber \\
\overline{m}^2_{\pi} & = & m^2 + \frac{\lambda}{2} \left[ \varphi^2 + \Delta_{\sigma} 
+ (N+1) \Delta_{\pi}\right]
\eea
The relation between 1PI and 2PPI expansion now simplyfies to :
\bea
\Gamma^{1PI} (m^2,\varphi) & = & S(\varphi) + \Gamma^{2PPI}_q (\overline{m}^2_{\sigma},
\overline{m}^2_{\pi},\varphi)  \\ \nonumber
& - & \frac{\lambda}{8} \int d^D x \left[ 3\Delta^2_{\sigma} + (N^2 - 1) \Delta^2_{\pi}
+ 2(N-1)\Delta_{\sigma} \Delta_{\pi}\right]
\eea
and the gapequations are :
\bea
\frac{\delta \Gamma^{2PPI}}{\delta \overline{m}^2_{\sigma}}
 & = & \frac{\Delta_{\sigma}}{2}
 \\ \nonumber
\frac{\delta \Gamma^{2PPI}}{\delta \overline{m}^2_{\pi}}
 & = & (N-1) \frac{\Delta_{\pi}}{2}
\eea
Two remarks are in order here. The derivation given above is independent of temperature,
so the relation (13) is also valid at finite T. Secondly, the masses $\overline{m}^2_{\sigma}$
and $\overline{m}^2_{\pi}$ are 2PPI effective masses. The physical $\sigma$ and $\pi$ 
masses $m^2_{\sigma}$ and $m^2_{\pi}$ still have to be calculated from the effective
action (as poles of the propagators) and are not identical to these 2PPI effective masses.

\section{Renormalisation of the 2PPI expansion}

To be useful for practical calculations, we have to show that equation (4) which relates
1PI and 2PPI expansions and the gapequations (8) can be renormalised with the conventional
counterterms. The crucial point in the proof of equations (4) and (8) was that the 2PPR 
insertions could be exactly summed via the effective 2PPI mass given in equation (3).
For this to remain true after renormalisation, we have to use a mass independent 
renormalisation scheme. Therefore, in this paper, we will use minimal subtraction.
Again, just as in the previous section, we will earmark the 1PI graphs by applying a
$\varphi$ derivative so that 2PPR and 2PPI parts are unambiguous.

We first renormalise the bubble subgraphs. Consider a generic bubble inserted at the
2PPR point $x$ (fig. 3.a). All primitively divergent subgraphs of the bubble graph
which do not contain the 2PPR point $x$ can be renormalised with the counterterms of
the $O(N)$ linear $\sigma$-model :
\bea
\delta \cal{L} & = & \delta Z \frac{1}{2} \partial_{\mu} \phi_i \partial_{\mu} \phi_j +
\frac{1}{2} \delta Z^{ij;k\ell}_2 m^2_{ij} \phi_k \phi_{\ell}  \\ \nonumber
& + & \frac{\lambda}{4!} \delta Z^{ijk\ell}_{\lambda} \phi_i \phi_j \phi_k \phi_{\ell}
\eea
where
\be
\delta Z^{ijk\ell}_{\lambda} = \delta Z_{\lambda} \left( \delta_{ij} \delta_{k\ell} +
\delta_{ik} \delta_{j\ell} + \delta_{i\ell} \delta_{jk}\right)
\ee
As a consequence of these subtractions, the contribution of the bubbles to the effective
mass is proportional to $\langle \phi_i \phi_j\rangle_c$ where the connected V.E.V. is
now calculated with the full Lagrangian, counterterms included. For subgraphs of the
bubble which do contain the 2PPR point $x$, we need only the
2PPR-parts of the counterterm. 
This means those parts which correspond to subtractions for subgraphs
which disconnect from the rest of the graph when two lines meeting at the 2PPR point $x$
are cut. Let's first renormalise the proper subgraphs of the bubble which contain $x$. Their 
generic topology is displayed in fig. 3.b and 3.c. They can be made finite with the
2PPR part $ \left( \frac{1}{2!}\right)^2 \delta Z^{ij;k\ell}_{\lambda,2PPR} \phi_i
\phi_j \phi_k \phi_{\ell}$ where the lines meeting at the 2PPR point $x$ carry the O(N)
indices i and j. Their contribution to the effective mas $m^2_{ij}$ is given by
$\frac{1}{2} \delta Z^{ij;k\ell}_{\lambda,2PPR} \langle \phi_k \phi_{\ell}\rangle_c$. We 
still have to subtract the overal divergences of the bubble graph. Their generic
topology is displayed in fig. 3.d and 3.e for coupling constant renormalisation and
fig. 3.f for mass renormalization. Again only the 2PPR parts of
the counterterm contributions
have to be included and the overall divergences contribute $\frac{\lambda}{2} 
\delta Z^{ij;k\ell}_{\lambda,2PPR} \varphi_k \varphi_{\ell} + \delta 
Z^{ij;k\ell}_{2,2PPR} m^2_{k\ell}$. Adding the various contributions coming from
renormalizing the bubble graphs, we find for the renormalised effective mass :
\bea
\overline{m}^2_{R,ij} & = & m^2_{ij} + \lambda \left[ \varphi_i \varphi_j + \Delta_{ij}
\right] + \frac{\lambda}{2} \left[ \varphi^2 + \Delta_{kk}\right] \delta_{ij}
\nonumber \\
& + & \delta Z^{ij;k\ell}_{2,2PPR} m^2_{k\ell} + \frac{\lambda}{2} \delta
Z^{ij;k\ell}_{\lambda,2PPR} \left[ \Delta_{k\ell} + \varphi_k \varphi_{\ell}\right]
\eea
where $\Delta_{ij} = \langle \phi_i \phi_j\rangle_c$ and the V.E.V. is calculated
with inclusion of the counterterms.

Because we use a mass independent renormalization scheme, the 2PPR part of coupling
constant renormalisation can be related to multiplicative mass renormalisation. Indeed
lets consider a generic diagram for mass renormalisation which is proportional to
$m^2_{pq}$ (fig. 4.a). On the other hand, lets consider in fig. 4.b a generic 2PPR
coupling constant renormalisation graph inserted at the 2PPR point $x$. The latter
can be gotten from the former by replacing the mass $m^2_{pq}$ by the coupling 
constant $\lambda_{pqij}$ and summing over p and q. Therefore we have:
\be
\lambda \delta Z^{ij;k\ell}_{\lambda,2PPR} = \lambda \left( \delta_{ij} \delta_{pq} +
\delta_{ip} \delta_{jq} + \delta_{iq} \delta_{jp}\right)
\delta Z^{pq;k\ell}_2
\ee
or
\be
\delta Z^{ij;k\ell}_{\lambda,2PPR} = \delta_{ij} \delta Z^{pp;k\ell}_2 + 2 \delta
Z^{ij;k\ell}_2
\ee
In an analogous way, we can relate the 2PPR part of multiplicative mass renormalisation
to vacuum energy renormalisation. In minimal subtraction, vacuum diagrams are
logaritmically divergent and proportional to $m^4$. Their divergences are cancelled
with the counterterm :
\be
\delta E_{vac} = \frac{1}{2} \overline{m}^2_{ij} \overline{m}^2_{k\ell} \delta
\zeta^{ij;k\ell}
\ee
A generic divergent vacuum graph proportional to $m^4$ is given in fig. 5.a. A generic
2PPR part of mass renormalisation is given in fig. 5.b. Just as in the previous case,
it is clear that
\be
\delta Z^{ij;k\ell}_{2,2PPR} = \lambda \left( \delta_{ij} \delta_{pq} + \delta_{ip}
\delta_{jq} + \delta_{iq} \delta_{jp}\right) \delta \zeta^{pq;k\ell}
\ee
or
\be
\delta Z^{ij;k\ell}_{2,2PPR} = \lambda \left( \delta_{ij} \delta \zeta^{pp;k\ell} + 
2 \delta \zeta^{ij;k\ell}\right)
\ee

Using (19) and (22) the renormalised effective mass 
given by (17) can be written as :
\bea
\overline{m}^2_{R,ij} = m^2_{ij} & + & \lambda \left[ Z^{ij,k\ell}_2 (\varphi_k \varphi_{\ell}
+ \Delta_{k\ell}) + 2 \delta \zeta^{ij;k\ell} m^2_{k\ell}\right]\nonumber \\
& + & \frac{\lambda}{2} \left[ Z^{pp;k\ell}_2 (\varphi_k \varphi_{\ell} + 
\Delta _{k\ell}) + 2 \delta \zeta^{pp;k\ell} m^2_{k\ell}\right]
\eea
where
\be
Z^{ij;k\ell}_2 = \frac{1}{2}\left( \delta_{ik}\delta_{j\ell} + \delta_{i\ell}
\delta_{jk}\right) + \delta Z^{ij;k\ell}_2
\ee
If we introduce the renormalised local composite operators
\bea
\langle \phi_i \phi_j\rangle_{c,R} = 
\Delta_{ij,R} = Z^{ij;k\ell}_2 \Delta_{k\ell} + \delta Z^{ij;k\ell}_2 \varphi_k
\varphi_{\ell} + 2 \delta \zeta^{ij;k\ell} m^2_{k\ell}
\eea
the renormalised effective mass finally becomes :
\bea
\overline{m}^2_{R,ij} = m^2_{ij} + \lambda \left( \varphi_i
\varphi_j + \Delta_{R,ij}\right) + 
\frac{\lambda}{2} \left( \varphi^2 + \Delta_{R,kk}\right) \delta_{ij}
\eea
From this equation, it follows that $\Delta_{R,ij}$ must be finite.

Once we have renormalised the bubble subgraphs, the $\varphi$ derivative of 
$\Gamma^{1PI}_q$ can be written as :
\be
\frac{\delta}{\delta \varphi_k} \Gamma^{1PI}_{q,BR} = \frac{\partial}
{\partial \varphi_k} \Gamma^{2PPI}_q (\overline{m}^2_R,\varphi) +
\left[ \lambda \varphi_k \delta_{ij} + \lambda \left( \delta_{ik} \varphi_j + 
\delta_{jk} \varphi_i\right)\right] \frac{\partial \Gamma^{2PPI}_q}
{\partial \overline{m}^2_{R,ij}} (\overline{m}^2_R,\varphi)
\ee
where BR stands for bubble renormalised. Because there is no
overlap, having renormalised the bubble subgraphs, we can now
renormalize the 2PPI remainder (which contains the earmarked
vertex). Let us first consider mass renormalization. A subgraph
$\gamma$ in the 2PPI remainder of $\frac{\delta}{\delta
\varphi_k} \Gamma^{1PI}_{q,BR}$ that needs mass renormalisation
can be made finite with a counterterm $\delta
Z^{ij;k\ell}_2(\gamma) m^2_{ij} \phi_k \phi_{\ell}/2$. However
for any such subgraph $\gamma$, there are subgraphs
$\gamma^{\prime}$ obtained from $\gamma$ by replacing the mass
$m^2_{ij}$ with a seagull or renormalised bubble. These
subgraphs require coupling constant renormalization which
entails a counterterm $\delta Z^{ij;k\ell}_{\lambda,2PPR}
\frac{\lambda}{2} (\varphi_i \varphi_j + \Delta_{R,ij})\phi_k
\phi_{\ell}/2$. 
Taking into account the identity (19) of renormalization
constants for mass renormalisation and 2PPR coupling constant
renormalization, the effective counterterm for the mass-type
divergent subgraphs adds up to :
\bea
\frac{1}{2} \delta Z^{ij;k\ell}_2 (\gamma)m^2_{ij} \phi_k
\phi_{\ell} & + & \frac{1}{2}\left( \delta_{ij} \delta
Z^{pp;k\ell}_2 (\gamma) + 2 \delta Z^{ij;k\ell}_2
(\gamma)\right) \frac{\lambda}{2} (\varphi_i \varphi_j +
\Delta_{R,ij})\phi_k \phi_{\ell} \nonumber \\
& = & \frac{1}{2} \delta Z^{ij;k\ell}_2 (\gamma) \left[ m^2_{ij}
+ \lambda [\varphi_i \varphi_j + \Delta_{R,ij}] +
\frac{\lambda}{2} (\varphi^2 + \Delta_{R,\ell \ell})
\delta_{ij}\right] \phi_k \phi_{\ell} \nonumber \\
& = & \frac{1}{2} \delta Z^{ij;k\ell}_2 (\gamma)
\overline{m}^2_{R,ij} \phi_k \phi_{\ell}
\eea
which is exactly what is needed for mass renormalization of
$\Gamma^{2PPI}_q (\overline{m}^2_R,\varphi)$ in the right hand
side of (27). The remaining divergent subgraphs need wave
function renormalization or are of the coupling constant
renormalization type that cannot be generated by inserting
seagulls or bubbles in mass-type divergent subgraphs. They are
made finite by counterterms independent of mass and hence are
the same for left and right hand sides of equation (27).
Therefore we can conclude that in a mass independent
renormalisation scheme, equation (4) can be renormalised with the
available counterterms as~:
\be
\frac{\delta}{\delta \varphi_k} \Gamma^{1PI}_{q,R} (m^2,\varphi)
= \frac{\partial}{\partial \varphi_k} \Gamma^{2PPI}_{q,R}
(\overline{m}^2_R, \varphi) + \left[ \lambda \varphi_k \delta_{ij} +
\lambda (\delta_{ik} \varphi_j + \delta_{jk} \varphi_i)\right] 
\frac{\partial \Gamma^{2PPI}_{q,R}}{\partial
\overline{m}^2_{R,ij}} 
\ee

To proceed, we have to renormalize the gapequations (8). Using
essentially the same arguments as in the previous paragraphs, we
find that
\be
\frac{\partial \Gamma^{1PI}_{q,R}}
{\partial m^2_{ij}} (m^2,\varphi) = \frac{\partial \Gamma^{2PPI}_{q,R}}
{\partial \overline{m}^2_{R,ij}} (\overline{m}^2_R,\varphi)
\ee
From the pathintegral, we readily obtain :
\bea
\frac{\partial \Gamma^{1PI}_{q,R}}{\partial m^2_{ij}} (m^2,\varphi) 
& = & \frac{1}{2} Z^{ij;k\ell}_2 (\varphi_k \varphi_{\ell} + \langle
\phi_k \phi_{\ell}\rangle_c) + \frac{\partial}{\partial
m^2_{ij}} \delta E_{vac} \nonumber \\
& = & \frac{1}{2} \left( \varphi_i \varphi_j + Z^{ij;k\ell}_2
\Delta_{k\ell} + \delta Z^{ij;k\ell}_2 \varphi_k \varphi_{\ell}
+ 2 \delta \zeta^{ij;k\ell} m^2_{k\ell}\right) \nonumber \\
& = & \frac{1}{2} (\varphi_i \varphi_j + \Delta_{R,ij})
\eea
where we used (20) and (25). Since $\Gamma^{1PI}_R$ is finite it
follows that $\Delta_{R,ij}$ is finite. This reconfirms our
analysis of bubble renormalization where from the finiteness of
the renormalised effective mass (eq.(26)), we concluded that
$\Delta_{R,ij}$ defined by (25) is finite. Using
$\frac{\partial}{\partial m^2_{ij}} \Gamma^{1PI}_R = \varphi_i
\varphi_j/2 + \frac{\partial}{\partial m^2_{ij}}
\Gamma^{1PI}_{q,R}$ and equations (30) and (31) we finally
obtain the renormalised gap equations :
\be
\frac{\Delta_{R,ij}}{2} = \frac{\partial
\Gamma^{2PPI}_{q,R}}{\partial \overline{m}^2_{R,ij}}
(\overline{m}^2_R,\varphi) 
\ee
As in the unrenormalised case, these gap equations can be used
to integrate (29) :
\be
\Gamma^{1PI}_R(m^2,\varphi) = S(\varphi) + \Gamma^{2PPI}_{q,R}
\left( \overline{m}^2_R, \varphi\right) - \frac{\lambda}{8} \int
d^D x \left[ (\Delta_{R,ii})^2 + (2\Delta_{R,ij})^2\right]
\ee

Our renormalised equation (33) together with the renormalised
gap equations (32) enable us to sum seagulls and bubble graphs
in such a way that perturbative renormalisability is preserved.
To renormalise $\Gamma^{1PI}$, it is sufficient to renormalize
$\Gamma^{2PPI}$ using a mass independent
renormalisation scheme such as MS, calculate the renormalised
local composite operators $\Delta_{R,ij}$ from the gap
equations, and substitute them back in (33). The advantage of
the 2PPI expansion is that with the same (or even less)
calculational effort as goes into the perturbative calculation
of $\Gamma^{1PI}$, the seagull and bubble graphs are summed
order by order. The gap equations are local and can easily be solved
numerically. 

The previous analysis was independent of temperature. Because
$\Gamma^{1PI}$ can be renormalised at finite T with the
counterterms at T = 0, the same goes through for
$\Gamma^{2PPI}$. Therefore, our renormalised equations (32) and
(33) are valid at finite T.

\section{Goldstone's theorem}

If we choose $m^2_{ij} = m^2 \delta_{ij}$, we can use of the
O(N) symmetry to define the renormalised effective masses
$m_{R,\sigma}$ and $m_{R,\pi}$ and renormalised composite
operators $\Delta_{R,\sigma}$ and $\Delta_{R,\pi}$ as :
\be
\overline{m}^2_{R,ij} = \frac{\varphi_i \varphi_j}{\varphi^2}
\overline{m}^2_{R,\sigma} + \left( \delta_{ij} - \frac{\varphi_i
\varphi_j}{\varphi^2}\right) \overline{m}^2_{R,\pi}
\ee
\be
\Delta_{R,ij} = \frac{\varphi_i \varphi_j}{\varphi^2}
\Delta_{R,\sigma} + \left( \delta_{ij} - \frac{\varphi_i
\varphi_j}{\varphi^2}\right) \Delta_{R,\pi}
\ee
so that equation (3) becomes :
\bea
\overline{m}^2_{R,\sigma} & = & m^2 + \frac{3\lambda}{2} \left[
\varphi^2 + \Delta_{R,\sigma} + \frac{N-1}{3}
\Delta_{R,\pi}\right] \\ \nonumber
\overline{m}^2_{R,\pi} & = & m^2 + \frac{\lambda}{2} \left[
\varphi^2 + \Delta_{R,\sigma} + (N+1) \Delta_{R,\pi}\right]
\eea
Because of O(N) symmetry, $\overline{m}^2_{R,\sigma},
\overline{m}^2_{R,\pi}, \Delta_{R,\sigma}$ and $\Delta_{R,\pi}$
are O(N) invariant functions of $\varphi_i$. The relation
between the renormalised 1PI and 2PPI expansion now simplifies
to :
\bea
\Gamma^{1PI} (m^2,\varphi) & = & S(\varphi) +
\Gamma^{2PPI}_{q,R} \left( \overline{m}^2_{R,\sigma},
m^2_{R,\pi},\varphi \right)  \nonumber \\
& - & \frac{\lambda}{8} \int d^D x \left[ 3 \Delta^2_{R,\sigma}
+ (N^2 - 1) \Delta^2_{R,\pi} + 2(N-1)\Delta_{R,\sigma}
\Delta_{R,\pi} \right]\nonumber \\
\mbox{}
\eea
and the gapequations become :
\bea
\frac{\delta \Gamma^{2PPI}_{q,R}}{\delta
\overline{m}^2_{R,\sigma}} & = & \frac{\Delta_{R,\sigma}}{2} 
\nonumber \\
\frac{\delta \Gamma^{2PPI}}{\delta \overline{m}^2_{R,\pi}} & =
& (N-1) \frac{\Delta_{R,\pi}}{2}
\eea

Since because of O(N) symmetry, the effective masses
$\overline{m}_{R,\sigma}$ and 
$\overline{m}_{R,\pi}$ and the composite operators
$\Delta_{R,\sigma}$ and $\Delta_{R,\pi}$ are O(N) invariant,
Goldstone's theorem must be obeyed at any loop order of the 2PPI
expansion. To check this explicitely, we should not make the
mistake of identifying the effective mass $\overline{m}_{R,\pi}$
with the real physical pion mass $m_{\pi}$, defined as the pole
in the pion propagator. This pole should occur at $p^2 = 0$ and
hence we can use the effective action at $p = 0$, i.e. the
effective 1PI potential. Using (37), the renormalised 1PI
effective potential becomes :
\bea
V^{1PI}_R (m^2,\varphi) & = & V(\varphi) + V^{2PPI}_{q,R} \left(
\overline{m}^2_{R,\sigma}, \overline{m}^2_{R,\pi}, \varphi^2
\right) \nonumber \\
& - & \frac{\lambda}{8} \left( 3 \Delta^2_{R,\sigma} + (N^2 - 1)
\Delta^2_{R,\pi} + 2(N-1) \Delta_{R,\sigma}
\Delta_{R,\pi}\right) 
\eea
Since $V^{2PPI}_{q,R}$ is O(N) invariant we can use the standard
argument to show that $\frac{\partial^2 V^{1PI}}{\partial
\varphi_i \partial \varphi_j}$ has N-1 zero eigenvalues at any
order of the 2PPI loop expansion. More explicitely we find from
(35), (36) and (38) that
\be
\frac{\partial V^{1PI}_R}{\partial \varphi_i} = \varphi_i \left(
m^2 + \frac{\lambda}{2} \varphi^2 + \frac{3\lambda}{2} \left(
\Delta_{R,\sigma} + \frac{N-1}{3} \Delta_{R,\pi}\right) + 2
\frac{\partial V^{2PPI}_{q,R}}{\partial \varphi^2} \right)
\ee
and
\bea
\frac{\partial^2 V^{1PI}_R}{\partial \varphi_i \partial
\varphi_j} 
& = & \delta_{ij} \left( m^2 + \frac{\lambda \varphi^2}{2} +
\frac{3\lambda}{2} \left( \Delta_{R,\sigma} + \frac{N-1}{3}
\Delta_{R,\pi} \right) + 2 \frac{\partial
V^{2PPI}_{q,R}}{\partial \varphi^2} \right) \nonumber \\
& + & \lambda \varphi_i \varphi_j \left[ 1 + 3 \frac{\partial
\Delta_{R,\sigma}} {\partial \varphi^2} + (N-1) \frac{\partial
\Delta_{R,\pi}} {\partial \varphi^2} + 4 \frac{\partial^2
V^{2PPI}_{q,R}} {(\partial \varphi^2)^2} \right]
\eea
So, we have N-1 massless particles if
\be
m^2_{\pi} = m^2 + \frac{\lambda}{2} \varphi^2 +
\frac{3\lambda}{2} \left( \Delta_{R,\sigma} + \frac{N-1}{3}
\Delta_{R,\pi}\right) + 2 \frac{\partial
V^{2PPI}_{q,R}}{\partial \varphi^2} = 0
\ee
Using (40) we conclude that the masslessness of the pions is
nothing else than the equation of motion in the case of
spontaneous symmetry breaking.

\section{The effective potential at finite temperature}

In this section, we will calculate the effective potential at
finite T using the 2PPI expansion at one loop. Since there are
(N-1) effective masses $\overline{m}_{R,\pi}$ and one mass
$\overline{m}_{R,\sigma}$ running in the one loop vacuum
diagram, we have
\be
V^{2PPI}_q \left( \overline{m}^2_{R,\sigma},
\overline{m}^2_{R,\pi},\varphi^2 \right) = \frac{1}{2} \sum \int
\ln (k^2 + \overline{m}^2_{R,\sigma}) + \frac{N-1}{2} \sum \int
\ln(k^2 + \overline{m}^2_{R,\pi})
\ee
where
\be
\sum \int = T \sum_{\omega_n} \int \frac{d^3 p}{(2\pi)^3}
\ee
and the Matsubara frequencies are denoted by $\omega_n$. We can
simply renormalise $V^{2PPI}_q$ using for example the
$\overline{MS}$ scheme and calculate the renormalised V.E.V. of
the composite operators from the gapequations (38). We find
at one loop :
\bea
V^{2PPI}_{q,R} & = & \frac{\overline{m}^4_{R,\sigma}}{64\pi^2}
\left( \ln \frac{\overline{m}^2_{R,\sigma}}{\overline{\mu}^2} -
\frac{3}{2} \right) + (N-1)
\frac{\overline{m}^4_{R,\pi}}{64\pi^2} \left( \ln
\frac{\overline{m}^2_{R,\pi}} {\overline{\mu}^2} -
\frac{3}{2}\right) \nonumber \\
& + & Q_T (\overline{m}_{R,\sigma}) + (N-1) Q_T
(\overline{m}_{R,\pi}) 
\eea
where
\be
Q_T(m) = T \int \frac{d^3 q}{(2\pi)^3} \ln \left( 1 -
e^{-\frac{\omega_q}{T}} \right)
\ee
The effective 1PI potential then reads :
\bea
V^{1PI}_R (m^2,\varphi) & = & \frac{m^2}{2} \varphi^2 +
\frac{\lambda}{8} \varphi^4 +
\frac{\overline{m}^4_{R,\sigma}}{64\pi^2} \left(\ln
\frac{\overline{m}^2_{R,\sigma}}{\overline{\mu}^2} - \frac{3}{2}\right) \nonumber
\\ 
& + & (N-1) \frac{\overline{m}^4_{R,\pi}}{64\pi^2} \left( \ln
\frac{\overline{m}^2_{R,\pi}}{\overline{\mu}^2} -
\frac{3}{2}\right) + Q_T(\overline{m}_{R,\sigma}) +
(N-1)Q_T(\overline{m}_{R,\pi}) \nonumber \\
& - & \frac{\lambda}{8} \left( 3 \Delta^2_{R,\sigma} + (N^2 - 1)
\Delta^2_{R,\pi} + 2(N-1) \Delta_{R,\sigma}
\Delta_{R,\pi}\right) 
\eea
with
\bea
\Delta_{R,\ast} & = & \sum \int \frac{1}{k^2 +
\overline{m}^2_{R,\ast}} \nonumber \\
& = & \frac{\overline{m}^2_{R,\ast}}{16\pi^2} \left( \ln
\frac{\overline{m}^2_{R,\ast}}{\overline{\mu}^2} - 1 \right) +
P_T (\overline{m}_{R,\ast})
\eea
where
\be
P_T(m) = 2 \frac{\partial}{\partial m^2} Q_T (m^2) = \int
\frac{d^3 q}{(2\pi)^3} \frac{n_B (\omega_q)}{\omega_q}
\ee

Our expression (47) together with the gapequations (48) and the
definition of the effective masses (36) completely agree with
previously published results [4,10] obtained using the CJT approach
at the daisy and superdaisy order (2PI expansion). The advantage
of our 2PPI expansion is that we arrive quite simply and
naturally at this result keeping only the one loop term while
the 2PI approach has to keep part of the 2 loop graphs (the
two bubble graph) and the simple expression (47) is only
obtained after some rearrangement. Furthermore, one can easily
calculate higher order terms in the 2PPI expansion while in the
2PI expansion, it is very difficult to go beyond the Hartree
approximation because of the non-locality of the gapequations.

In our approach, renormalisation of the non-perturbative results
is straightforward. This is because we renormalise the effective
2PPI potential and hence the gapequations before we try solving
them. If one does it the other way around as in [4],
perturbative renormalizability is apparently spoiled. This is
because in the resummation (see section 3), parts of the
counterterms (the 2PPR parts) have to be included at all orders,
and it is very difficult if not impossible to do this once the
gapequations are solved and whole classes of diagrams have
already been summed.

As to our numerical results, they coincide (at least for that
part concerning the effective potential) with the CJT results
obtained for example in [10]. We take the N=4 Gell-Mann Levy
linear $\sigma$-model, relevant for QCD and use the parameter
choice of Nemoto et al. [10]: $\lambda = \frac{90.2}{3}$ (our
$\lambda$ differs from the one in [10] by a factor of 3 at N=4)
$\overline{\mu} = 320$ $\mbox{MeV}$, $m^2 = - 122375 \mbox{MeV}^2$. Our results
are in the chiral limit. Extension to real pion masses is
trivial. In fig. 6, we display the effective potential at T =
186, 192, 200 and 208 $\mbox{MeV}$. We clearly see a first order phase
transition around $T_c = 200 \mbox{MeV}$. This agrees with other mean
field approaches [15,16,5]. The renormalisation
group however, leads us to believe that the actual phase
transition of the O(4) linear sigma model should be second
order. There are suggestions [10] that inclusion of the 2 loop
setting sun diagram should change the phase transition from
first to second order, at least for small $\lambda$. However this
is very difficult to check in the CJT formalism because of the
non-local nature of the gap equations. In our 2PPI approach,
this should be no problem and work concerning this issue is in
progress [17].

\section{The $\sigma$-meson mass}

Because of its relevance in the context of ultrarelativistic
heavy-ion collisions, the $\sigma$-resonance has been thoroughly
studied in various models [18,19,6,10].
In the CJT approach to the O(4) linear
$\sigma$-model, the $\sigma$-meson mass has been studied in [10]
and defined via the effective potential. The physical
$\sigma$-meson however, is defined via the effective action as
solution of the mass equation
\be
\int d^4 x e^{ip(x-y)} \frac{\delta^2 \Gamma}{\delta
\varphi_j(y) \delta \varphi_i(x)} = 0
\ee
for $-p^2$. At finite temperature, the propagators are no longer
Lorentz invariant. They are functions of $p^2_0$ and
$\vec{p}^2$ instead of $p^2$. The standard prescription
is to define mass at rest with respect to the heatbath, this
means putting $\vec{p}^2 = 0$ and solving (50) for $- p^2_0$.
Using the fact that at one loop the 2PPI effective action only
depends on $\varphi_i$ through the effective masses, it follows
from (37) and the gapequations (38) that
\bea
\frac{\delta^2 \Gamma}{\delta \varphi_j(y) \delta \varphi_i(x)}
& = & \left[ - \partial^2 + m^2 + \frac{\lambda}{2}
\varphi^2 (x) + \frac{3\lambda}{2} \left(
\Delta_{R,\sigma}(x) + \frac{N-1}{3}
\Delta_{R,\pi}(x)\right) \delta_{ij}\right. \nonumber \\ 
& + & \left. \lambda \varphi_i(x)
\varphi_j(y) \right] \delta (x-y) 
+ \frac{3\lambda}{2} \varphi_i(x) \left[ \frac{\delta
\Delta_{R,\sigma}(x)} {\delta \varphi_j(y)} + \frac{N-1}{3}
\frac{\delta \Delta_{\pi}(x)}{\delta \varphi_j(y)}\right]
\nonumber \\
& & \mbox{}
\eea
If we choose $\varphi_i = \delta_{iN} \varphi$ and use equation
(36) for $\overline{m}^2_{R,\sigma}$, we can rewrite the
$\sigma$-massequation as :
\be
p^2 + \overline{m}^2_{R,\sigma} + 3\lambda \varphi^2
\left( \Delta_{R,\sigma}^{\prime} (p) + \frac{N-1}{3}
\Delta^{\prime}_{R,\pi} (p)\right) = 0
\ee
with
\be
\Delta^{\prime}_{R,\ast}(p) = \int d^4 x e^{ip(x-y)}
\frac{\partial \Delta_{R,\ast}(x)}{\partial \varphi^2 (y)}
\ee
From the equation of motion (42) at one loop and equation (36)
it follows that $\overline{m}^2_{R,\sigma} = \lambda \varphi^2$
which is the tree-level mass of the $\sigma$-meson (the
condensate $\varphi$ is of course determined by the full
expression (42) containing quantum corrections). Therefore the
self energy of the $\sigma$-meson at one loop in the 2PPI
expansion is given by :
\be
\Sigma_{\sigma}(p) = 3\lambda \varphi^2 \left(
\Delta^{\prime}_{R,\sigma} (p) + \frac{N-1}{3}
\Delta^{\prime}_{R,\pi} (p)\right)
\ee

This selfenergy can be calculated exactly. From equation (48) we
have
\be
\Delta^{\prime}_{R,\ast}(p) = - B_{\ast} \frac{\partial
\overline{m}^2_{R,\ast}} {\partial \varphi^2}
\ee
where
\be
B_{\ast} = B(\overline{m}^2_{R,\ast},p) = \Sigma \int \frac{1}{q^2
+ \overline{m}^2_{R,\ast}}\frac{1}{(q+p)^2 +
\overline{m}^2_{R,\ast}} 
\ee
From (55) and the effective mass equations (36) we derive the
system 
\bea
\Delta^{\prime}_{R,\sigma} & = & - B_{\sigma} \left(
\frac{3\lambda}{2} + \frac{3\lambda}{2}
\Delta^{\prime}_{R,\sigma} + \frac{\lambda}{2} (N-1)
\Delta^{\prime}_{R,\pi}\right) \nonumber \\
\Delta^{\prime}_{R,\pi} & = & - B_{\pi} \left( \frac{\lambda}{2}
+ \frac{\lambda}{2} \Delta^{\prime}_{R,\sigma} +
\frac{\lambda}{2} (N+1) \Delta^{\prime}_{R,\pi}\right)
\eea
This system can be easily solved and we finally obtain for the
self energy of the $\sigma$-meson :
\be
\Sigma_{\sigma}(p) = - \lambda^2 \varphi^2 \frac{9B_{\sigma} +
(N-1) B_{\pi} + 3\lambda (N+2) B_{\sigma} B_{\pi}}{2 + 3\lambda
B_{\sigma} + \lambda (N+1) B_{\pi} + \lambda^2 (N+1) B_{\sigma}
B_{\pi}} 
\ee
Adding the effective mass $\overline{m}^2_{R,\sigma}$ 
(which runs in the tree level propagators of the 2PPI expansion)
to the one loop 2PPI selfenerg (58), we find we have summed an
infinite series of Feynman diagrams given in fig. 7. The
propagators in the internal lines are $\sigma$ as well as
$\pi$-propagators and they carry effective masses
$\overline{m}_{R,\sigma}$ and $\overline{m}_{R,\pi}$. This sum
goes beyond the daisy-superdaisy approximation which is given by
the first term only. In fact we have summed all 2PPR
contributions to the selfenergy which can be made from one loop
2PPI subdiagrams. This is of course consistent with the fact
that we have calculated the $\sigma$-meson propagators from the
one loop 2PPI effective action.

In the same way, we can calculate the one loop 2PPI mass of the
pion. Our one loop 2PPI approximation again goes beyond the
daisy-superdaisy result and we find that the selfenergy is just
enough to make the pion mass exactly zero. This is of course
easily understood as the effective action at p = 0 is nothing else
than the effective potential and we have already shown on
general grounds in section 4 that the second derivative of the
effective potential with respect to the pion fields is zero at
the minimum of the potential.

To obtain numerical results we have to evaluate
$B(\overline{m}^2_{R,\ast},p)$ at finite T. Using dimensional
regularisation and the $\overline{MS}$ scheme, we find
\bea
B_R(m^2,p) = - \frac{1}{16\pi^2} \left[ \ln
\frac{m^2}{\overline{\mu}^2} + \sqrt{1 + \frac{4m^2}{p^2}}
\ln \left( \frac{\sqrt{1 + \frac{4m^2}{p^2}} + 1}
{\sqrt{ 1 + \frac{4m^2}{p^2}} - 1} \right) - 2 \right] +
B_T(m^2,p)\nonumber \\
\mbox{}
\eea
with
\bea
B_T(m^2,p) & = & \int \frac{d^3 q}{(2\pi)^3} n_B (\omega_{\bf q})
\frac{1}{\omega_{\bf q}} \frac{\omega^2_{\bf {q+p}} -
\omega^2_{\bf q} + p^2_0}
{(\omega^2_{\bf{q+p}} - \omega^2_{\bf q} + p^2_0)^2 +
4\omega^2_{\bf q} p^2_0} \nonumber \\
& + & n_B(\omega_{\bf{q+p}}) \frac{1}{\omega_{\bf q+p}} \frac
{- \omega^2_{\bf{q+p}} + \omega^2_{\bf q} + p^2_0}
{(- \omega^2_{\bf{q+p}} + \omega^2_{\bf q} + p^2_0)^2 +
4\omega^2_{\bf{q+p}} p^2_0}
\eea
We determine the $\sigma$-meson mass as the zero in $p^2_0
(\overline{p}^2 = 0)$ of the real part of the inverse
$\sigma$-propagator $p^2 + \overline{m}^2_{R,\sigma} +
Re(\Sigma_{\sigma} (p^2_0, \vec{p}^2 = 0))$. We again use the
parameters $\lambda = 90.2/3,
\overline{\mu} = 320 \mbox{MeV}, m^2 = - 122375 \mbox{MeV}^2$. In Nemoto et al
[10], the $\sigma$-meson mass was determined from the effective
potential and the parameters were chosen such that $m_{\sigma} =
600 \mbox{MeV}$ at T = 0. Our more physical definition of the
$\sigma$-mass gives $m_{\sigma} = 548.3 \mbox{MeV}$ at T = 0. So the
correct definition of mass only gives a 10\% change and
therefore, this choice of parameters is acceptable given the
ambiguity in the experimental value for the $\sigma$-meson mass.
In fig. 8 we display the physical $\sigma$-meson mass (zero in
$p^2_0$ of $p^2_0 + \overline{m}^2_{R,\sigma} + Re
(\Sigma_{\sigma}(p_0, \vec{0}))$ and the $\sigma$-meson mass as
determined from the effective potential and equal to
$\overline{m}^2_{R,\sigma} + \Sigma_{\sigma}(0,\vec{0})$, at
finite temperature. The influence of temperature is to decrease
to $\sigma$-meson mass, a well known effect established with
other methods [6] or in other models [18,19]. 

\section{Summary and conclusions}

In this paper, we have studied the O(N) linear $\sigma$-model at
finite temperature using the 2PPI expansion. We have shown that
at one loop order in this expansion, the Hartree result is
reproduced in a very efficient way. We have given an all orders
proof that this expansion can be renormalised with the usual
counterterms if a mass independent renormalisation scheme is
used. We have shown that at finite temperature and each order of
the 2PPI expansion, Goldstone's theorem is obeyed. We have
calculated the effective potential for N =4 and found a first
order phase transition as was to be expected from a mean field
approximation. However, whereas previous methods to obtain the
Hartree result such as the CJT formalism are very difficult to
apply beyond the mean field level, the 2PPI expansion is ideally
suited to investigate post-Hartree corrections to various
thermodynamical quantities. For example, we calculated the one
loop 2PPI result for the $\sigma$-meson mass and showed that it
sums an infinite series of diagrams which go beyond the daisy
and superdaisy approximation (Hartree approximation). As to the
thermodynamics of the phase transition, higher order loops in the
2PPI expansion can give important corrections which could change
the order of the phase transition. Work on the 2 loop correction
is in progress [17]. Also, one of us (H. Verschelde) showed in [20]
, that the 2PPI expansion can be renormalisation group
resummed. The combination of the 2PPI expansion and the
renormalisation group may in the future lead to a better picture
of the thermodynamics of the O(N) linear $\sigma$-model.

%----------------------------------figure captions------------------------------
\newpage
\noindent
{\bf Figure captions}\\

\noindent Fig1: Generic 2PPR diagram\\

\noindent Fig2: 2PPR part is shaded, 2PPI rest is earmarked\\ 

\noindent Fig3: Generic bubble(shaded) and its subdivergences(shaded).Thick lines are full 
propagators.\\

\noindent Fig4: Generic diagram for mass renormalisation (a) and generic 2PPR coupling constant
renormalisation diagram (b)\\ 

\noindent Fig5: Generic divergent vacuum diagram (a) and generic 2PPR mass
renormalisation diagram (b)\\

\noindent Fig6: Effective potential $V(\varphi)$ at $T=186,192,200,208$ for $\lambda 
=\frac{90.2}{3}$.\\

\noindent Fig7: Feynman diagrams contributing to $\sigma$ - propagator at one loop in the 2PPI 
expansion.\\

\noindent Fig8: Two ways to determine the sigma mass in function of the temperature. Upper 
line: 
using the second derivative of the effective potential, lower line: using the second 
variational derivative of the effective action.

%----------------------------------figures--------------------------------------
\newpage
\begin{figure}
\centering
\includegraphics[width=10cm,height=10cm]{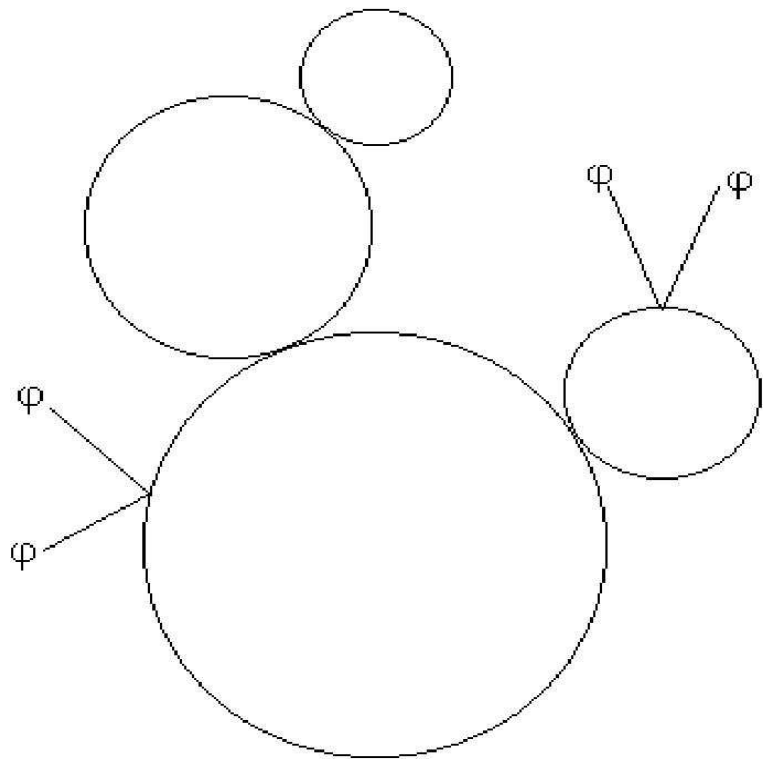}
\caption{Generic 2PPR diagram}
\end{figure}
\newpage
\begin{figure}
\centering
\includegraphics[width=10cm,height=12cm]{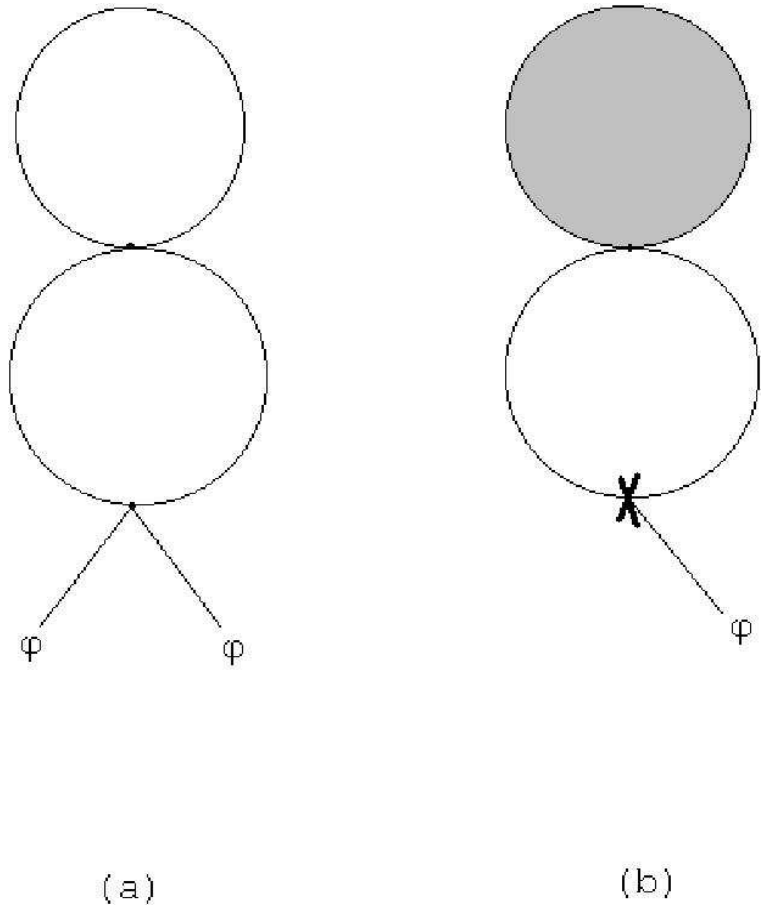}
\caption{2PPR part is shaded, 2PPI rest is earmarked}
\end{figure}
\newpage
\begin{figure}
\centering
\includegraphics[width=14cm,height=20cm]{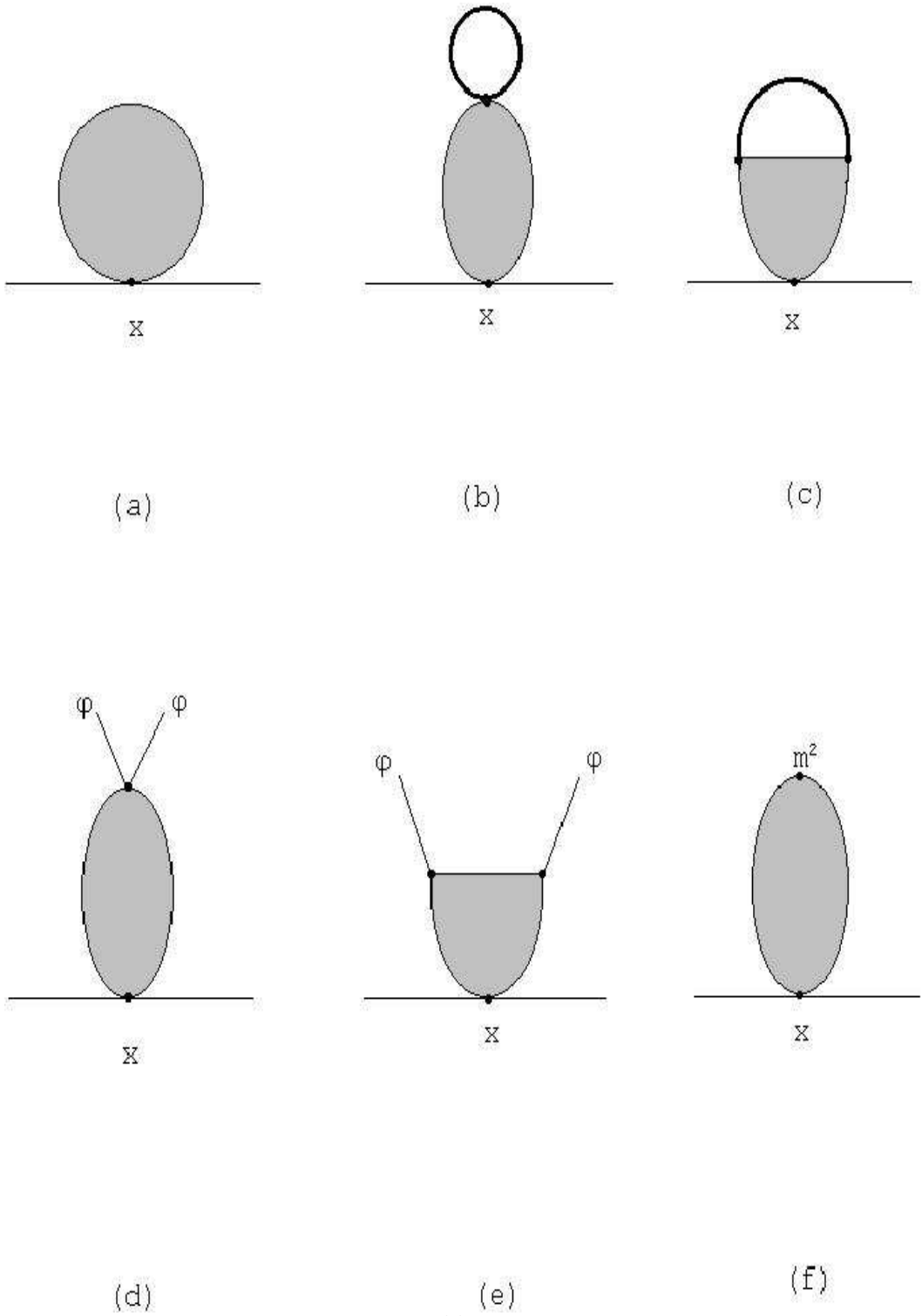}
\caption{Generic bubble(shaded) and its subdivergences(shaded).Thick lines are full 
propagators.}
\end{figure}
\newpage
\begin{figure}
\centering
\includegraphics[width=10cm,height=12cm]{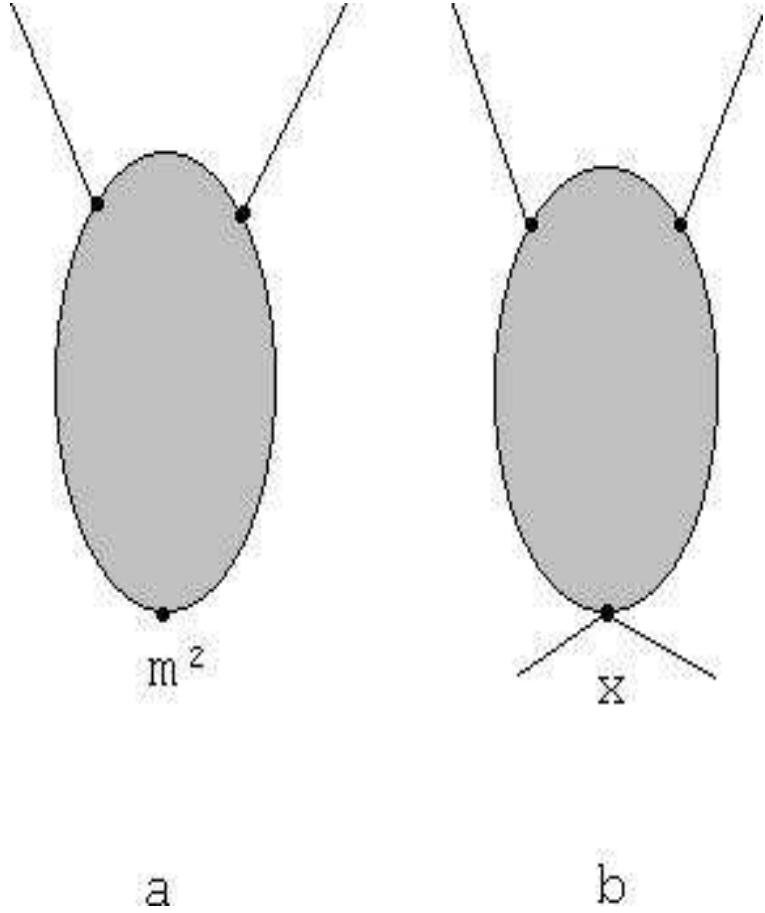}
\caption{Generic diagram for mass renormalisation (a) and generic 2PPR coupling constant
renormalisation diagram (b)}
\end{figure}
\newpage
\begin{figure}
\centering
\includegraphics[width=10cm,height=12cm]{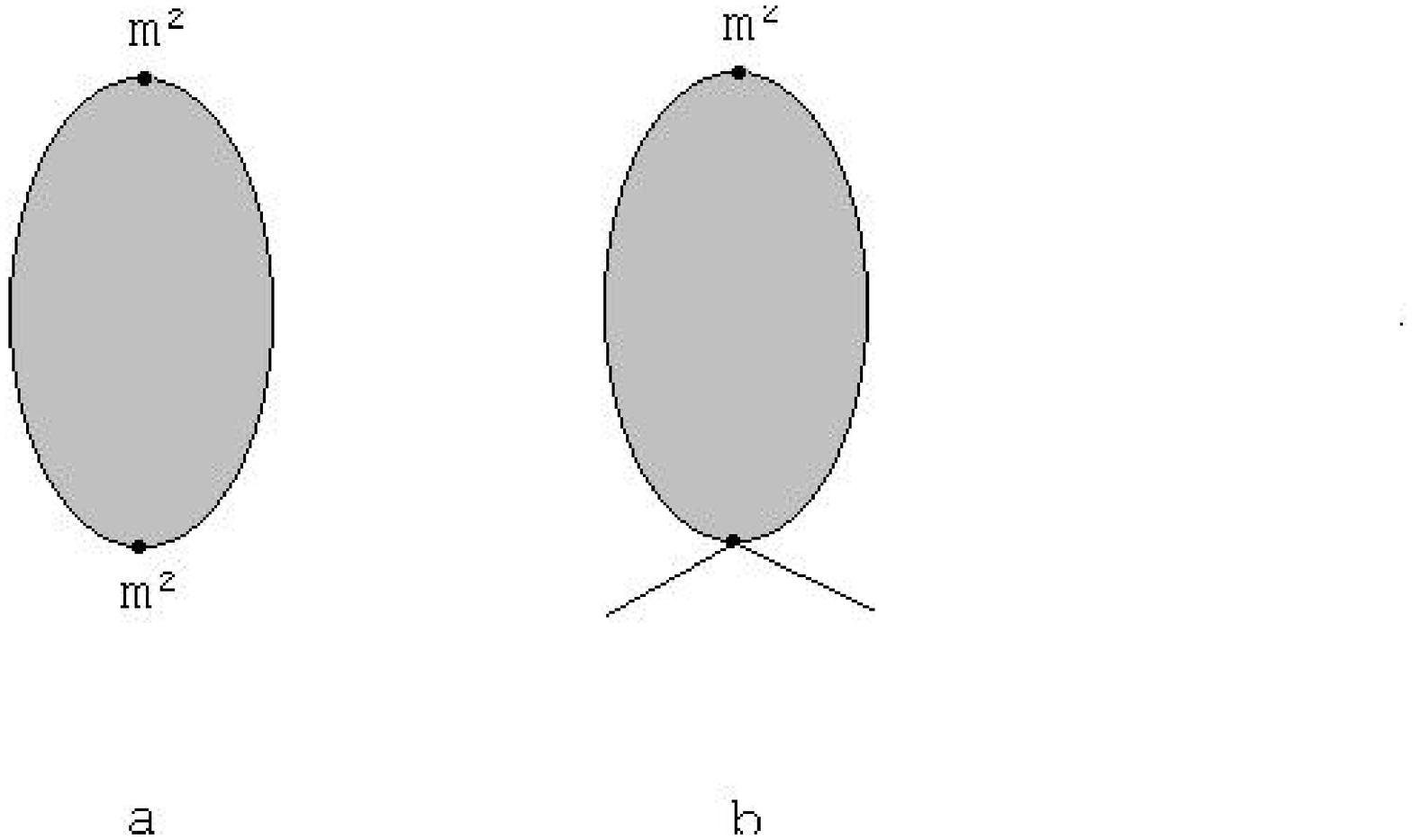}
\caption{Generic divergent vacuum diagram (a) and generic 2PPR mass
renormalisation diagram (b)}
\end{figure}
\newpage
\begin{figure}
\centering
\includegraphics[width=12cm,height=10cm]{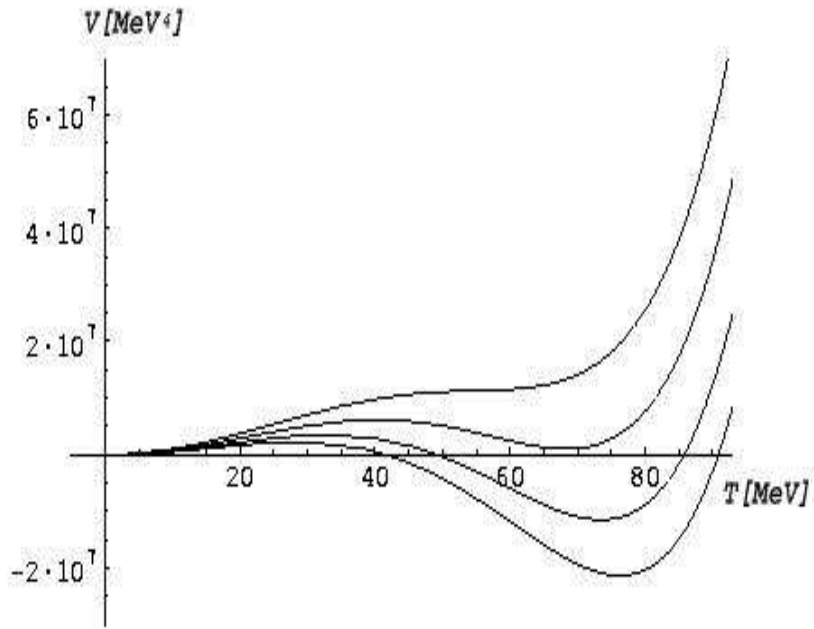}
\caption{Effective potential $V(\varphi)$
at $T=186,192,200,208$ for $\lambda 
=\frac{90.2}{3}$.}
\end{figure}
\newpage
\begin{figure}
\centering
\includegraphics[width=14cm,height=4cm]{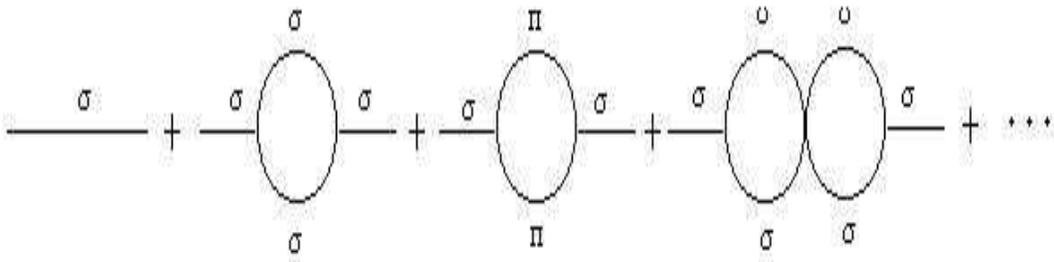}
\caption{Feynman diagrams contributing to $\sigma$ - propagator at one loop in the 2PPI 
expansion}
\end{figure}
\newpage
\begin{figure}
\centering
\includegraphics[width=12cm,height = 10 cm]{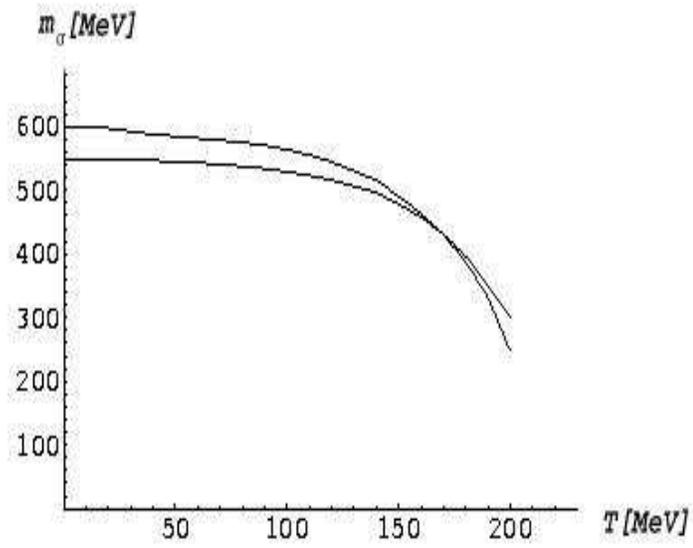}
\caption{Two ways to determine the sigma mass in function of the 
temperature. Upper line: 
using the second derivative of the effective potential, lower line: using 
the second 
variational derivative of the effective action.}
\end{figure}

\end{document}